\documentclass{JINST}

\title{Sensitivity of an image plate system in the XUV (60 eV \textless \ E 
\textless \ 900 eV)}

\author{
{B. H. Failor}$^{a}$\thanks{ Corresponding author.}{, E. M. 
Gullikson}$^{b}${, N. G. Link}$^{a}${, J. C. 
Riordan}$^{a}${, and B. C. Wilson}$^{c}$\\
\llap{$^a$}L-3 Applied Technologies, Pulse Sciences, Inc.,\\
  2700 Merced St., San Leandro, CA 94577, USA\\
\llap{$^b$}Lawrence Berkeley National Laboratory,\\
  1 Cyclotron Road, Berkeley, CA 94720, USA\\ 
\llap{$^c$}Defense Threat Reduction Agency,\\
  Fort Belvoir, VA 22060-6201, USA\\ 
  E-mail: \email{bruce.failor@l-3com.com}}

\abstract{Phosphor imaging plates (IPs) have been calibrated and 
proven useful for quantitative x-ray imaging in the 1 to over 1000 keV 
energy range. In this paper we report on calibration measurements made at 
XUV energies in the 60 to 900 eV energy range using beamline 6.3.2 at the 
Advanced Light Source at Lawrence Berkeley National Laboratory. We measured 
a sensitivity of \textasciitilde 25 $\pm$ 15 counts/pJ over the stated 
energy range which is compatible with the sensitivity of Si photodiodes that 
are used for time-resolved measurements. Our measurements at 900 eV are 
consistent with the measurements made by Meadowcroft \textit{et al}. at \textasciitilde 1 
keV.}


\begin{document}

\section{Introduction}

Extreme Ultraviolet (XUV) photons are difficult to image because they 
interact strongly with most materials. Back illuminated charge-coupled 
device (CCD) cameras have been used [1], but are 
limited in active area, are difficult to clean, and the hardware is 
expensive. XUV sensitive film, such as Kodak 101, which does not have an 
overcoat, is fragile and requires chemical processing. The Computed 
Radiography (CR) phosphor imaging plate (IP) is an intriguing alternative 
because they have a wide number of applications, from medical radiography 
(the application that drove their initial development) to plasma 
spectroscopy research [2] due to their (1) large 
linear dynamic range, \textasciitilde 10$^{5}$, (2) low cost, 
\textless {\$}100, (3) availability in large areas, \textasciitilde 30 cm, 
and (4) sensitivity which is compatible with Si photodiodes. Some imaging 
plates have been engineered to efficiently detect beta-particles from 
tritium. One such plate is the ``Kodak Storage Phosphor Screen TR'' (Ref 
1041540). Meadowcroft \textit{et al}. have previously reported on the sensitivity and 
fading characteristics of IPs for 1-75 keV x 
rays.[3] In this paper we present sensitivity 
measurements for the ``Kodak Storage Phosphor Screen TR'' plate in the 60 - 
900 eV energy range, demonstrating that IPs can be successfully used for XUV 
imaging. 

\section{Technical Approach}

The IP contains phosphor that can store a latent image which is later 
digitized by raster scanning its surface with a laser beam and recording the 
luminescence with a photomultiplier tube (PMT).[3] 
We used a plate which had a BaFBr phosphor, activated with Eu, chemical 
formula BaFBr(Eu). The phosphor was deposited on a flexible plastic 
substrate and the surface facing the XUV source was protected with a thin 
layer of cellulose acetate.[4] The IP was digitized 
using a commercially available scanner,[5] and 
software,[6] which allowed the following parameters 
to be set: (1) scan speed, (2) spatial resolution, and (3) PMT voltage 
(which determines the PMT gain). PMT voltage directly affects the digitized 
image pixel value amplitudes and the spatial resolution specifies the length 
and width of the square pixels. Scan speed determines how rapidly the image 
is scanned and a slower speed will result in a larger dynamic range because 
there is more time for an efficient transfer of the stored energy on the 
plate. The latent image stored on the plate does fade with 
time,[3] so the time between exposure and 
digitization needs to be accounted for in order to achieve the highest 
accuracy possible.

We exposed the IP to XUV radiation at the Advanced Light Source (ALS) at 
Lawrence Berkeley National Laboratory. Beamline 6.3.2 provides virtually 
monochromatic (E/$\Delta $E $\ge $ 7000) photons in the energy range of 
interest, 50 - 1000 eV.[7] The exposure chamber 
includes calibrated detectors that allow the XUV flux to be monitored, 
shutters, apertures and filters for modulating the beam intensity, and a 
sample stage for positioning the object under test, in our case the IP, in 
and out of the photon beam.

\section{Experiments and results}

Two days of IP exposures will be described here. The goal of the first day 
was to get measurements of the IP sensitivity for just a few energies spread 
over a wide range. During the second day we focused on relatively closely 
space energies at low photon energy (E \textless 100 eV) and also at the C 
and O K-edges where we expected to see modulations in sensitivity due to 
absorption in the cellulose acetate protective layer.

\begin{figure}[h!]
\begin{center}
\centerline{\includegraphics[width=5.83in,height=4.92in]{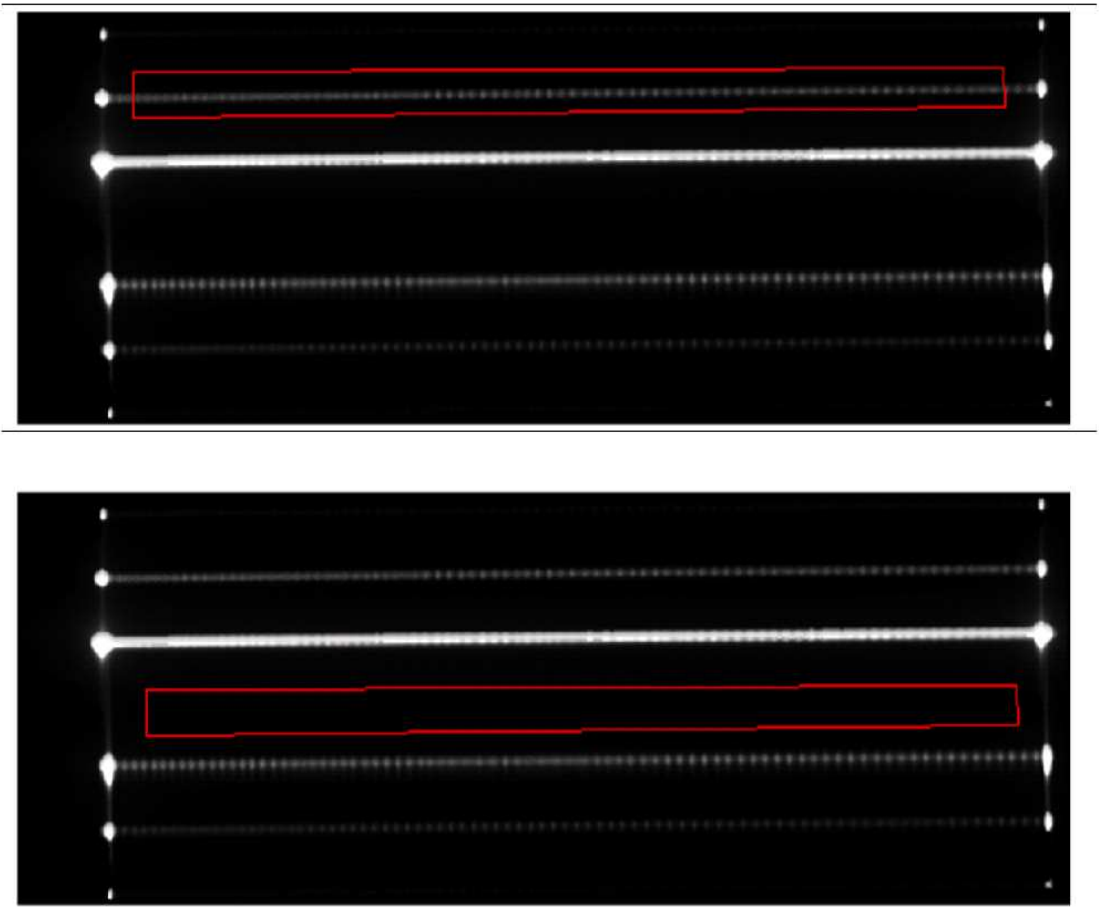}} \par \
\caption{During the first day a number of different diameter apertures were used to 
produce a range of exposures on the IP. Top: Region of interest (ROI) over 
which the image was integrated is outlined in red. This region corresponds 
to a photon energy of 185 eV and a beamline aperture of 3 mm. Bottom: ROI 
corresponds to no exposure and will be used for background subtraction. The 
exposures from top to bottom in this image were separated by 3 mm on the IP 
and correspond to (1) 185 eV, 1 mm aperture, (2) 185 eV, 3 mm aperture, (3) 
185 eV, no aperture (\textasciitilde 10 mm diameter opening), (4) no 
exposure (provides background level), (5) 400 eV, no aperture 
(\textasciitilde 10 mm diameter opening), (6) 400 eV, 3 mm aperture, (7) 400 
eV, 1 mm aperture. For the weakest two scans, (1) and (7), only the starting 
and ending points, where the XUV beam rested between scans, are clearly 
visible on the images. The scans look nonuniform (like a series of dots) 
because the stepping mechanism used to position the IP did not move at a constant velocity.}
\label{fig1}
\end{center}
\end{figure}

On the first day, two IP exposures were done, one IP at 900 eV followed by a 
second IP with exposures taken at 100, 185, 400, and 500 eV. By changing a 
beamline aperture, power levels between \textasciitilde 0.1 and 
\textasciitilde 100 nW were achieved. Since the XUV spot size was only 
\textasciitilde 100 $\mu$m vertically by \textasciitilde 300 $\mu$m 
horizontally, the IP was scanned at an average speed of 0.75 cm/sec in the 
vertical direction to produce regions of relatively uniform exposure with 
peak fluences in the 5-5000 nJ/cm$^{2}$ range (corresponding to 0.5-500 
pJ/pixel for 100 $\mu$m square pixels). A range of exposures were used 
initially because the sensitivity of the image plate to the XUV was unknown. 
An example of an IP image from the first day is shown in 
\figurename~\ref{fig1}. The IP scanner settings were 600 V PMT 
voltage, and a 4000 rpm scan speed. The spatial resolution was held at 100 
microns for all the scans reported here. Because the IPs were scanned within 
5 minutes of exposure, compensation for fading was not 
necessary.[3]

On the second day of experiments we captured two images. The first exposure 
explored the low energy range and included exposures at 60, 65, 75, 80, and 
100 eV in the vicinity of the Ba M4 and M5 edges. The second exposure was 
done at 270, 280, 290, 300, 530, 540, 550, and 560 eV, bracketing the C K 
and O K edges. For this second day, the scan speed was reduced to 2000 rpm. 
We think that this changes the shape of the exposed lines slightly, but not 
the integrated values that are reported here. In contrast to the first day, 
the IPs were not scanned within 5 minutes of exposure, but rather not until 
125 to 197 minutes afterwards. Previous experimental measurements 
[3] for a phosphor without an overcoat indicate a 
fading time constant of \textasciitilde 36 min while a phosphor with an 8 
micron overcoat had a measured time constant of \textasciitilde 57 min. As 
shown below, x-ray measurements indicate that the overcoat thickness for the 
IP described here is well below 1 micron, but even if the longer time 
constant is used, the fact that the fading component amplitude is just 
31{\%} of the total signal [3] means that the 
maximum error that could be introduced is from 0.8 to 2.5{\%}. Because of 
its small amplitude, we will neglect the fading component in the discussion 
that follows. The measured signals for the two IPs exposed on the second day 
should be 69{\%} of the amplitude that would have been measured if the IP 
had been scanned within a few minutes of the exposure, rather than 2-3 hours 
later.

\begin{figure}[h!]
\begin{center}
\centerline{\includegraphics[width=5.9in,height=4.21in]{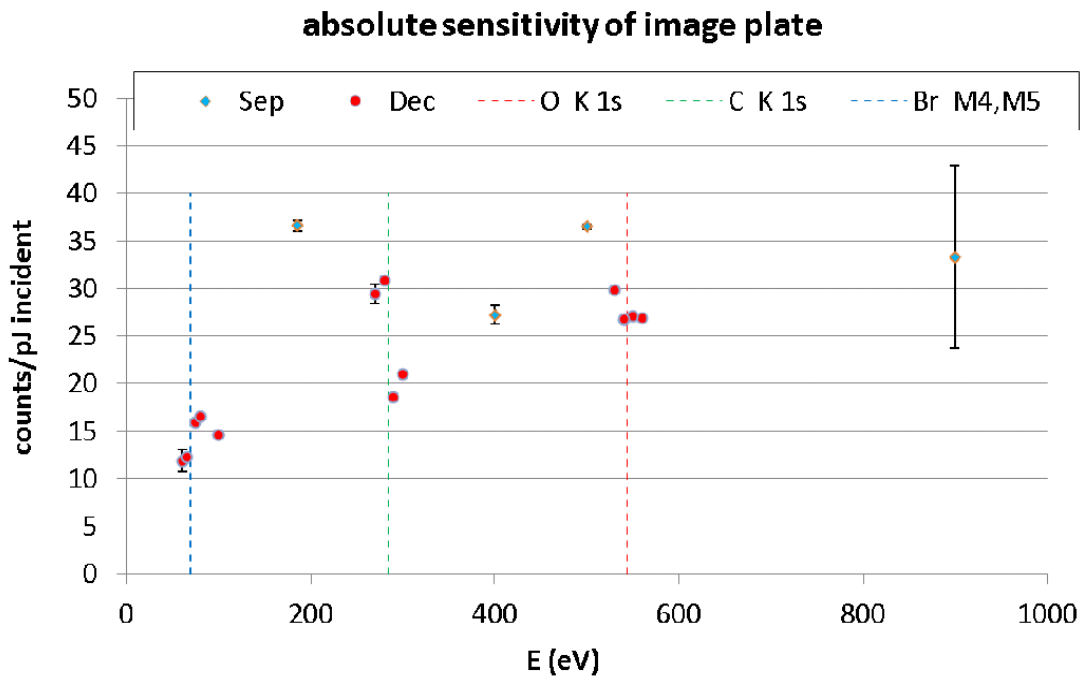}} \par \
\caption{The sensitivity of an image plate-image plate scanner combination is \textasciitilde 25 $\pm$ 15 counts/pJ over the energy range 60-900 eV.  The sensitivity is expected to drop at the O and C K-edges since both elements are present in the cellulose acetate overcoat, and increase at the Br M4 and M5 edges because Br is present in the phosphor.}
\label{fig2}
\end{center}
\end{figure}

The results for all the sensitivity measurements are given in 
\tablename~\ref{tabl1} and \figurename~\ref{fig2}. 
Because absolute intensity measurements were available for the first day, 
and only relative intensity measurements for the second due to a recording 
error, the second day data is normalized to the first at 100 eV, the only 
energy common to both days. The reported values were found by (1) 
integrating an area on the image corresponding to a particular energy 
exposure, (2) subtracting the integral of the same area in an unexposed 
region of the image (see example in \figurename~\ref{fig1}), and 
(3) normalizing by the product of the length of the integral on the IP and 
the energy deposited per unit length by the ALS.

The effective thickness of the cellulose acetate protective layer can be 
estimated from the height of the absorption feature corresponding to the C K 
edge, assuming a slow variation in the phosphor response for the energies 
270, 280, 290, and 300 eV. The chemical formula for the cellulose acetate 
was not specified, so estimates were made for 
C$_{6}$H$_{2}$O$_{2}$(OH)$_{3}$ and 
C$_{6}$H$_{2}$O$_{2}$(OOCCH$_{3})_{3}$., the two extremes of 
C$_{6}$H$_{2}$O$_{2}$(OH)$_{3-m}$(OOCCH$_{3})_{m}$ , 0 $\le $m $\le $ 3. 
The areal densities found from the fits of the natural logarithm of signal 
versus linear mass absorption coefficient are given in 
Table 2. The areal density value estimate, 
\textasciitilde 20 $\mu$g/cm$^{2}$, is about 50 times lower than the 
expected thickness based on an application of \textasciitilde 1 mg/cm$^{2}$ 
of coating material. At least 3 factors could influence this result. First, 
the phosphor particles as deposited on the IP backing along with a binder, 
will form a surface with finite roughness and voids that would wick up the 
overcoat material, which we assume is applied as a liquid. Secondly, the 
phosphor particles nearer the surface of the IP would have a higher 
probability of contributing to the detected signal because they have a 
higher probability of (1) being stimulated by the readout laser and (2) 
having their luminescence escape to the IP surface. Finally, when XUV 
photons produce electrons in the overcoat, phosphor, and binder, the 
phosphor particles near the surface are most likely to be excited by those 
electrons and the secondary electrons that they produce.

The sensitivity of \textasciitilde 25 counts/pJ is consistent with the 
values reported by Meadowcroft \textit{et al}. for a different IP system. In the low 
energy limit they reported sensitivities \textasciitilde 0.5 
``mPSL/keV.''[3] If ``mPSL'' is taken to be 0.001 
count, the corresponding sensitivity would be 3.4 counts/pJ, less than a 
factor of 10 different than what we have reported here. We consider this 
acceptable agreement for two IP systems employing similar technology but 
produced by different suppliers and operated by two different research 
groups.

\section{Conclusions}

We have measured the sensitivity of an image plate at XUV energies between 
60 eV and 900 eV and the results look promising for imaging applications. 
The absorption features in the cellulose acetate overcoat only modulate the 
sensitivity by \textasciitilde 60{\%} of the peak value, and the sensitivity 
at 60 eV is better than 30{\%} of the peak value. The sensitivity of the IP, 
25 $\pm$ 15 counts/pJ for a 600 V PMT bias and 100 micron square pixels, 
is compatible with 0.1 A (5 V signals recorded by a digitizer with 50 ohm 
termination), 20 ns FWHM signals measured from 1 mm square photodiodes with 
a sensitivity of 0.2 A/W. The corresponding energy fluence, 1 
$\mu$J/cm$^{2}$, would produce a signal of 2500 counts out of a possible 
\textasciitilde 50,000, if the IP were scanned with 100 micron resolution. 
From the 600 V PMT setting used here, the bias can be varied between 300 and 
1200 V to accommodate changes in the XUV source brightness.

\acknowledgments

The authors gratefully acknowledge useful discussions with S. A. Mango, 
Carestream NTD, who also provided the IP that was used in this study. The 
Department of Energy provided the ALS beamtime and support for E. M. 
Gullikson. The remainder of the work was sponsored by the Defense Threat 
Reduction Agency (cleared for public release by the Defense Threat Reduction 
Agency, dated 25 January 2012, reference number NT-12-077 [PA-12-058]).

\newpage 

\newpage 

\newpage 

\begin{table}
\begin{center}
\caption{Sensitivity data were acquired on two different days.  On the first day, the data was absolutely calibrated.  Only a relative calibration was obtained on the second day, so the sensitivity measurements for that day have been scaled at the photon energy common to both days - 100 eV.}
    \begin{tabular}{|c|c|c|}
        \hline
        calibration & photon E (eV) & sensitivity (counts/pJ) \\ \hline\hline
        absolute    & 100           & 14.60                   \\ \hline
        absolute    & 185           & 36.58 $\pm$\ 0.56       \\ \hline
        absolute    & 400           & 27.23 $\pm$\ 0.94       \\ \hline
        absolute    & 500           & 36.50 $\pm$\ 0.28       \\ \hline
        absolute    & 900           & 33.31 $\pm$\ 9.61       \\ \hline\hline
        relative    & 60            & 11.90 $\pm$\ 1.17       \\ \hline
        relative    & 65            & 12.30                   \\ \hline
        relative    & 75            & 15.94                   \\ \hline
        relative    & 80            & 16.50                   \\ \hline
        relative    & 270           & 29.40 $\pm$\ 1.03       \\ \hline
        relative    & 280           & 30.87                   \\ \hline
        relative    & 290           & 18.53                   \\ \hline
        relative    & 300           & 20.99                   \\ \hline
        relative    & 530           & 29.85                   \\ \hline
        relative    & 540           & 26.73                   \\ \hline
        relative    & 550           & 27.04                   \\ \hline
        relative    & 560           & 26.88                   \\ 
        \hline
    \end{tabular}
\label{tabl1}
\end{center}
\end{table}

\newpage 

\begin{table}
\begin{center}
\caption{Fits of the IP sensitivity near the C K-edge provide and estimate of the cellulose acetate areal density, assuming that the signal variation is dominated by the change in XUV absorption.  Fit parameters, r2 and F, are given in the table as well.}
    \begin{tabular}{|c|c|c|c|c|}
        \hline
        cellulose acetate formula & areal density ($\mu$g/cm$^{2}$) & unattenuated sensitivity (counts/pJ) & r2    & F  \\ \hline\hline
        C$_{6}$H$_{2}$O$_{2}$(OH)$_{3}$        & 21.0 $\pm$\ 2.3                 & 32.9 $\pm$\ 1.3                       & 0.977 & 84 \\ \hline
        C$_{6}$H$_{2}$O$_{2}$(OOCCH$_{3})_{3}$  & 18.8 $\pm$\ 2.1                 & 32.4 $\pm$\ 1.3                       & 0.976 & 81 \\ \hline\hline
        average                   & 19.9 $\pm$\ 2.2                 & 32.7 $\pm$\ 1.3                       & -     & -  \\ 
        \hline
    \end{tabular}
\label{tab2}
\end{center}
\end{table}


\begin{thebibliography}{9}

\bibitem{bib1}
B. H. Failor, N. Qi, J. S. Levine, H. Sze, and E. M. Gullikson,
{\emph{Rev.\ Sci.\ Instrum.} {\bf 75} (2004) 4026}.

\bibitem{bib2}
D. B. Sinars, D. F. Wenger, S. A. Pikuz, \textit{et al}., 
{\emph{Rev.\ Sci.\ Instrum.} {\bf 82} (2011) 063113}.

\bibitem{bib3}
A. L. Meadowcroft, C. D. Bentley, and E. N. Stott, 
{\emph{Rev.\ Sci.\ Instrum.} {\bf 79} (2008) 113102}.

\bibitem{bib4}
S. A. Mango, private communication.

\bibitem{bib5}
Scan X Discover HC (Serial {\#}1010), ALLPRO Imaging.

\bibitem{bib6}
``PixelRay'' Image Acquisition and Analysis Software, L-3 Communications.

\bibitem{bib7}
J. H. Underwood, E. M. Gullikson, M. Koike, P. J. Batson, P. E. Denham, K. D. Franck, R. E. Tackaberry, and W. F. Steele, 
{\emph{Rev.\ Sci.\ Instrum.} {\bf 67} (1996) 3372}.

\bibitem{bib8}
B. H. Failor, H. M. Sze, J. W. Banister \textit{et al.}, 
{\emph{Physics of Plasmas} {\bf 14} (2007) 022703}.

\end{thebibliography}
\end{document}